# The Maintenance of Sex: Ronald Fisher meets the Red Queen

David Green[1], Chris Mason[1]

[1]Department of Anatomy, University of Otago Medical School, Great King Street, Dunedin, New Zealand

Running head: Maintenance of sex

Corresponding author:

David Green
Department of Anatomy
University of Otago Medical School
Great King Street
Dunedin 9016
New Zealand
e-mail: david.green@anatomy.otago.ac.nz
phone: (+64) (3) 479 7439
fax: (+64) (3) 479 7254

**ABSTRACT**

**Background:** Sex in higher diploids carries a two-fold cost of males that should reduce its fitness relative to cloning and result in its extinction. Instead, sex is widespread and it is clonal species that face early obsolescence. One possible reason is that sex is an adaptation to resist ubiquitous parasites, which evolve rapidly and potentially antagonistically.

**Results:** We use a heuristic approach to model mutation-selection in finite populations where a parasitic haploid mounts a negative frequency-dependent attack on a diploid host. The host evolves reflexively to reduce parasitic load. Both host and parasite populations generate novel alleles by mutation and have access to large allele spaces. Sex outcompetes cloning by two overlapping mechanisms. First, sexual diploids adopt advantageous homozygous mutations more rapidly than clonal diploids under conditions of lag load. This rate advantage can offset the lesser fecundity of sex. Second, a relative advantage to sex emerges under host mutation rates that are fast enough to retain fitness in a rapidly mutating parasite environment and increase host polymorphism. Clonal polymorphic populations disproportionately experience interference with selection at high mutation rates, both between and within loci. This slows clonal population adaptation to a changing parasite environment and reduces clonal population fitness relative to sex. The interference increases markedly with the number of loci under independent selection. Rates of parasite mutation exist that not only allow sex to survive despite the two-fold cost of males but which enable sexual and clonal populations to have equal fitness and co-exist.

**Conclusion:** We develop a heuristic model of mutation and selection involving a diploid host in a parasite environment. The approach establishes two conditions under which sex is supported: one creates different rates of adoption of advantageous mutations; the other creates differences in selection efficiency in removal of deleterious mutations. Since all organisms carry parasitic loads, the model is of general applicability.

**INTRODUCTION**

Sexual reproduction in higher eukaryotes is an evolved adaptation that uses a well-defined mechanism for mixing DNA in progeny (segregation, meiosis, gametogenesis and syngamy). One feature of this reproductive adaptation is the emergence of males, which impose costs on sex that are not borne by clonal reproduction [1-4]. Cloning should be widely prevalent as a consequence and sex should be under constant threat of extinction. The reverse is the case.

The costs of sex were identified in two ways. Maynard Smith [1] and Williams [2] independently identified a cost of meiosis associated with the production of haploid gametes by random sampling of diploid genomes. Although fusion of haploid gametes reinstates diploidicity in the offspring, individual offspring are now no longer copies of their parents. Maynard Smith also identified a fecundity problem: the two-fold cost of males [3]. Clonal mothers produce twice as many reproducing mothers in each generation as sexual mothers because the sexual mothers also produce males. If males and females are produced in equal numbers, the clonal mother and her progeny should rapidly drive the sexual population to extinction.

Maynard Smith's challenge [3] is greater still if a sexually-reproducing population is to resist invasion by a sexual mother converting to clonal reproduction because, at the moment of conversion, the adaptation of the new clonal mother is the same as her former sexual self. Her fitness therefore doubles. If she survives drift, clonal invasion of the sexual population is rapid and sex is extinguished. However, invasion could be prevented if the clonal mother and her offspring suffered a rapid decline in fitness relative to sex, followed by fitness-dependent loss of clones from the population and their extinction.

The cost of meiosis and the two-fold cost of males are not the same thing. A sexual population that is genetically homogeneous will have no cost of meiosis because there can be no loss of genetic relatedness between parents and offspring. However, conversion of a sexual mother to clonality under these conditions will still extinguish sex, because of the fitness difference. There are at least three known differences between clonal and sexual populations that might account for differences in fitness under otherwise similar circumstances.

First, there is a hypothetical difference in the rate of adoption of advantageous mutations by sexual and clonal populations (abbreviated to the KJ effect) [5]. An advantageous mutation in a sexual population sweeps uninterruptedly from one homozygous state to a new, better-adapted homozygous state in a single, seamless transition, whereas a clone is arrested at the heterozygote stage. It requires a second advantageous mutation to proceed. Because the target size for the second mutation is half that of the first mutation, it takes more than twice as long, on average, for clones to move from one homozygous state to the next. Sex should show an ongoing rate advantage over cloning if the external environment changes constantly to create opportunities for continuous adoption of new advantageous mutations. This is a moving optimum for the diploid population. It is an environment that was foreshadowed by Felsenstein [6] ("gene substitutions act to counteract the effects of environmental change, as if the population were running as hard as it could to stay in the same place."[6]). Felsenstein's paper also introduced the concept of lag load as a measure of the mismatch between an organism and its environment. The approach adopted by Kirkpatrick and Jenkins [5] was analytical, and relied on a number of simplifying assumptions. These were subsequently examined in some detail [7] and brought the earlier conclusions [5] into doubt. The advantage to sex also disappears when mutation rates are low, and mitotic recombination becomes an alternative explanation for the improved adaptation of clones [8]. Taken at their face value, these studies [7,8] appear largely to have excluded a rate advantage to sex in adopting beneficial mutations and minimized the importance of the KJ effect.

Fisher [9] had earlier identified two different sources of advantage to sex. In the first, two beneficial mutations occurring near-contemporaneously in separate individuals can rapidly be brought onto the same chromosome by sexual reproduction and crossing-over, whereas a clonal population must adopt the two beneficial mutations sequentially. This delays the adjustment of clonal populations to maladaptation. Muller [10] independently stated the same idea, and it is now known as the Fisher-Muller hypothesis. It requires mutations of small effect, large populations and high mutation rates for sex to acquire its advantage [11,12]. Fisher also identified a disadvantage to clones that is, in effect, a source of selective interference [9]. In polymorphic clonal populations, the polymorphism is reflected as a series of reproductively isolated clonal lines, any one of which can receive an advantageous mutation. If a clonal line is poorly adapted, the advantageous mutation may be lost as the polyclonal population moves to monoclonality through expansion of a fitter clone, whereas in a sexual population, an advantageous mutation can switch its background through independent assortment and crossing-over. Poorly-adapted alleles in clonal populations may also be difficult to remove efficiently. A similar argument to Fisher's was proposed by Manning [13]. Muller also suggested that clonal interference, where clones with similar fitness have to compete, would blunt selection on clones [10], something that has been supported subsequently both theoretically and experimentally [14]. No party appears to have shown that the lower adaptation of clonal populations, by whatever mechanism, is sufficient to reduce the fitness of these populations below that of equivalent sexual populations, allowing sex to succeed. However, the elimination of interference between coupled loci by crossing-over is widely seen as conferring a major fitness advantage on sex [15] because it allows elimination of Hill-Robertson interference.

Both the KJ effect [5] and the Fisher effect [9] can arguably be considered within the wider context of mutation-selection theory, even though Fisher's argument was not couched in mathematical terms. Both mechanisms provide a nominal advantage to sex in circumstances where there is acute relief of populations from maladaptation. Intermittent advantages are unlikely to produce consolidation of a major adaptation such as sex, which is likely to require, instead, the near-endless generation of opportunities, either for differential adoption of advantageous mutations, or for the widespread existence of polymorphism at key loci. It was Hamilton and his colleagues [16,17] who suggested that sex was an adaptation to resist parasites, because parasites were frequently pathogenic and capable of endless, rapid evolution. However, Hamilton's models were based on recombination and selection, not mutation and selection. Like many others, they assumed that sex secures its advantage through crossing-over, not independent assortment. They also used a haploid model in which the cost of sex was taken as the cost of meiosis. Their model cannot exploit the KJ effect that potentially operates in diploid populations [5], nor can it generate the full range of inter-allelic and inter-haplotypic interference in rapidly changing environments that is also open to diploids.

At its simplest, mutation and selection in a static environment may be taken to reflect the competing claims of stabilizing selection, which reduces genetic variation, and new mutations that tend to increase it. Mathematical treatment remains challenging, even for a static environment [18], but is more so when selection varies with time [19-23]. Given the historic difficulty in developing analytical methods for dealing with changing environments, we chose to employ a heuristic approach. This paper describes implementation of a computer simulation to examine the adoption of mutations in a target diploid species exposed to evolving parasites exerting negative frequency-dependent selection.

**RESULTS**

***Recapitulating Kirkpatrick and Jenkins***

We confirm the KJ effect [5] by computer simulation. In the simplest case, where the parasite is a uniform population that is held fixed, the simulation provides a 17-point numerical scale to measure the adaptive value of an individual host allele. Diploid genotypes have a 33-point scale because of the averaging of allelic adaptations. Fig. 1(a) shows the response of a uniform diploid population that is maladapted by one mutation per allele in a static environment. The starting adaptation score of the diploid genotype is 1.0 - 0.0625 = 0.9375. The sexual population moves to an average adaptation score of 1.0 after adoption of one beneficial mutation, whereas the clonal population moves to an average adaptation score of 0.96875 after the first mutation, and 1.0 after the second. Fig. 1(a) is the same as Fig. 1 in [5] but generated computationally in real time using our simulation. Adoption of beneficial mutations in the simulation is stochastic. The average time to adoption in the sexual population (measured at population adaptation = 0.953) = 231 generations (Fig. 1 (b)). Fig. 1(c) shows the same simulation for a clonal population. The average time to adoption (measured at population adaptation = 0.953) = 228 generations, whereas average time to adoption of the second mutation (measured at 0.984) is 681 generations. Adjusting for the time from first adoption, the time to adoption of the second mutation is close to 3X that of the first. A ratio of ~3 is expected since the target size for the second mutation is half the size of that for the first mutation and the clock starts for the second mutation as the first mutation is adopted. This behaviour is general across a range of mutation rates and population sizes.

***The KJ effect with multiple loci***

Kirkpatrick and Jenkins [5] adopted an analytical approach to the calculation of population fitness for sexual and clonal populations over the medium term, and made a number of simplifying assumptions. These included: (i) loci were numerous; (ii) loci were binary, i.e. 0 or 1; and (iii) loci fully adopted mutations before another locus proceeded to adoption. These, and other assumptions, are restrictive and have been noted as such [8]. We avoid the restrictions by setting mutation rates independently for hosts and parasites and allowing for stochastic adoption of mutations. We then compute all adaptation scores on-the-fly for each allele for each generation, placing no limitations on the order of mutation adoption. We also study single loci and small numbers of loci where the parasite environment evolves continuously to reduce the adaptation of the host through negative frequency-dependent selection. A key relationship at steady state is the ratio of mutation rate in the parasite, $\mu_p$, to that in the host, $\mu_h$, since a function of this ratio gives the lag load, $LL$ (i.e. $LL = f(\mu_p/\mu_h)$). As noted in the Introduction, a lag load provides opportunities for continuous adoption of advantageous mutations by the host [6,3].

Data are shown in Figures 2(a-d) for a range of lag loads generated in two populations ($N = 100, 3000$) as parasite mutation rates increase at fixed host mutation rates (Figures 2(a,c), $\mu_h = 10^{-6}$ bits/allele/generation; Figures 2(b,d), $\mu_h = 10^{-3}$ bits/allele/generation). Adaptation scores decline as parasite mutation rates increase relative to those of the host (i.e. $\mu_p/\mu_h$ increases). The average adaptation scores of the sexual populations (black curves) are always higher than that of the clonal populations (red curves) under equivalent conditions.

The long-run steady state values of population adaptation that arise at low host mutation rates ($10^{-5}$ or less) are largely independent of the absolute values of the mutation rates $\mu_p$ and $\mu_h$, but are a function of the ratio, $\mu_p/\mu_h$. At these host mutation rates, sexual populations are essentially monomorphic (effective allele number = ~ 1.0), and clones are dimorphic (effective allele number = ~2.0), for reasons discussed shortly. Maladapted alleles are cleared efficiently. What is seen in Figures 2(a) and (c), therefore, is a straight Kirkpatrick and Jenkins (KJ) advantage to sex emerging as lag loads increase (that is, ratio of effective allele number, clonal/sexual tends to 2). Large lag loads are needed for the effect because these loads indicate maladaptation, and that, in turn, increases the target size for adoption of advantageous mutations at a particular locus. This dilutes the chances that an advantageous mutation will convert a clonally heterozygote position to homozygosity, because of the number of alternative targets, and this secures the kinetic advantage to sex. The progressive loss of lag load with better adaptation of the host sees a reduction in the size of the mutation target and an increase in the chance of heterozygotes converting to homozygotes (reflected in trend of effective allele number to 1). When host matching of the environment reaches a maximum, and lag loads are small, there is little or no kinetic advantage to sex in adopting advantageous mutations and the lower fecundity of sexual reproduction reduces its fitness relative to clones. The data in Figure 2(c) show an adaptation advantage to sex of ~2.3 when $\mu_h = 10^{-6}$ and $\mu_p = 10^{-4}$, which attenuates as $\mu_p$ progressively declines. Figures 2(b) and (d) show a second effect emerging as host mutation rates are increased. For a host mutation rate in Figure 2(d) of $10^{-3}$, the parasite mutation rates needed to produce the same lag loads as $\mu_p$ at $10^{-4}$ in Figure 2(c) are $10^{-2}$ for sex and $\sim 3 \times 10^{-3}$ for clones, an $\sim$10-fold reduction in $\mu_p/\mu_h$ for sex and $\sim$35-fold reduction of $\mu_p/\mu_h$ in clones. At the same time, the number of alleles rises to 18.9 for the sexual population and 20.2 for the clonal, from 1.00 and 2.01 respectively when $\mu_h = 10^{-6}$ under similar lag loads. This large increase in polymorphism with increase in host mutation rate interferes with clearance of maladapted alleles, even in sexual populations, but the effect is larger in clonal populations. We return to this point in presenting the data that underpins Figures 6(a-d).

So far, we have shown data for adaptation scores. Fitness, in this simulation, is a multiplicative function of adaptation and fecundity, and the two-fold cost of males in sexual reproduction is reflected as a two-fold fecundity difference between clonal and sexual populations. The green curves in Figures 2(b) and (d) represent the fitness of sexual populations relative to clonal populations obtained by halving the adaptation scores of sexual populations; clonal fitness is represented by the solid adaptation curve in red, since the fecundity factor is 1.0 relative to 0.5 for sexual reproduction. These data show that sexual populations are less fit than clonal ones at small lag loads, but a transition point exists past which sexual fitness exceeds clonal fitness. This transition point represents a parasite mutation rate, $\mu_p$, at which, for a given host mutation rate, $\mu_h$, the parasite environment causes an equal loss of fitness in both populations; above this parasite mutation rate, clonal populations are less fit than sexual ones. Importantly, this cross-over point sits within a not-implausible range of lag loads. When the same approach is adopted for data in Figure 2(a), there is effectively no cross-over point, and for data in Figure 2(c), it applies at relatively high lag loads. We conclude from these data that the KJ effect at low mutation rates, both $\mu_p$ and $\mu_h$, allows sexual populations to go a considerable way in matching the fitness advantage of clonal populations, subject to the existence of lag loads. The effect attenuates as these loads disappear.

### Emerging adaptation differences: understanding the mechanism

Data in the previous section showed differences in average adaptation scores of sexual (black curves) and clonal populations (red curves) under lag loads. Here we provide data that show how these effects arise. First, we extract information from the simulation about allele lifetimes and allele frequencies during the course of a simulation. For technical reasons related to the tractability of computation, these data are derived from diploid genotypes with three loci, each of 2 x 16 bits. Figures 3(a) and (b) show typical profiles for sexual and clonal populations, respectively, under middle-range lag loads (adaptation scores for individual loci: sex, 0.707; clone, 0.548) using mutation rates, $\mu_h$, per locus of $10^{-4}$ bits/allele/generation for hosts, and $10^{-2}$ bits/allele/generation for parasites; $N$ = 3000. Sexual reproduction at these mutation settings is characterized by a series of short-lived allele sweeps that mostly move towards temporary homozygosity (effective allele number = ~2). Occasionally, the population supports two alleles simultaneously. Lowering mutation rates without shifting the ratio, $\mu_p/\mu_h$, between parasite and host extends allele lifetime and effective allele number tends to 1.0 (data not shown). In all cases, alleles come into existence and then die, rather than cycle. This behaviour reflects the size of the allele landscape being used. The non-cycling behaviour of these alleles is in contrast to those cited in [17].

Clonal populations show significantly different allele behaviour under the same conditions. Here, maximum allele frequencies cap themselves at 0.5, as lag load develops and effective allele number tends to 2.0. As with sexual reproduction, lowering mutation rates without shifting the ratio, $\mu_p/\mu_h$, extends allele lifetime without lifting the frequency cap. We argue the explanation for the emergence of the cap is as follows. An average clonal adaptation score of 0.55 per locus (see previous paragraph) represents a mismatch of approximately 14 bits for a diploid locus of 2 x 16 = 32bits. If the one-bit components of the locus are homozygous (either 0,0 or 1,1) and adopt an advantageous mutation, they move to heterozygosity (0,1 or 1,0). The next advantageous mutation to undergo adoption by the locus has no better than a 1-in-14 chance of occurring at the heterozygotic bit in question. The chance of adoption is actually lower than that, as we show shortly, because alleles start losing adaptation as soon as they have been adopted. Taken together, the low chance of adopting a second mutation at the heterozygote locus and the deterioration in the adaptation of its carrying allele make its chances of moving to homozygosity effectively zero, and this is borne out by the data. The cap on the 0.5 frequency rises as the parasite mutation rate is progressively lowered (through $10^{-3}$, $10^{-4}$, $10^{-5}$, etc.) relative to the host (that is, as the value of $\mu_p/\mu_h$ falls, because the average adaptation score of the clonal population increases, as does the chance of adopting a heterozygote-converting advantageous mutation.

Histograms of allele frequency distribution are shown in Figure 4. At the low $\mu_p$ of $10^{-6}$ bits/allele/generation, with $\mu_h = 10^{-4}$ bits/allele/generation, the allele distributions of sexual and clonal populations are essentially identical, and both populations are almost entirely homozygotic (that is, effective allele number is close to 1.0). This reflects the effect of stabilizing selection in a constant environment. Sex has approximately half the fitness of cloning because of its lower fecundity. When the environment changes continuously, the average adaptation of the population falls. The precise level of the fall depends on the nature of the change. A random walk has almost no effect, whereas negative frequency-dependent selection imposes directional selection whose strength depends on the ratio $\mu_p/\mu_h$ and the mode of reproduction (sexual or clonal). Where significant lag loads emerge ($\mu_p$ of $10^{-3}$ in this particular case), there is capping of the highest clonal frequency. This result is an inevitable consequence of the KJ effect under significant lag load. It has the important effect of enforcing greater polymorphism (strictly, a higher effective allele number) on the clonal population than the sexual population, thereby creating a more substantial opportunity for negative interactions between co-alleles at single loci. The effect is independent of absolute mutation rates, but rests on a suitable value of $\mu_p/\mu_h$, which determines the lag load.

### *Allele lifetimes and declines in adaptation scores*

Here we present the data that provide evidence of greater relative interference with selection in clones as the number of loci increase. These data fall into two sections. First, we compare average allele lifetimes in sexual and clonal populations, where there are either one or three loci in contention (Figures 5(a-d). The data for the sexual populations show little difference between the distribution of allele lifetimes at a single locus (Figure 5(a)) and three loci (Figure 5(b)). Selection is marginally faster at the single locus, with an earlier time-to-peak, and peak frequency is slightly lower, reflecting the faster clearance of alleles with poor adaptation. This is also reflected as a faster fall in adaptation score. Figure 5(c) shows the average allele lifetime behaviour for a single clonal locus under the same conditions. The maximum allele frequency is much less, due to the capping effect at significant lag loads, and there is already evidence of intra-locus interference with selection, as evidenced by the long tail. Loci remain heterozygotic, establishing selective interference between different alleles at the same locus. The timeline of the adaptation score shows a rapid fall after the peak allele frequency is attained, increasing the likelihood that an advantageous mutation will occur in an allele of declining adaptation, with a co-allele whose adaptation is also declining. These data support the existence of intra-locus interference, even at single loci.

Figure 5 (d) shows the effects of coupling in a clonal three-locus genotype, which greatly extends the average lifetime of alleles and increases the effective allele number. The long tail is due to a combination of intra-locus and inter-locus interference with selection. The obligatory coupling that occurs in clones greatly increases the probability that a beneficial mutation will occur in a poorly adapted background. The coupling effect increases with the number of loci that are coupled, because overall effective allele number increases as a power function of the effective allele number at single loci, as well as contributing additional interference. The existence of multiple loci has relatively little effect in sexual reproduction if the loci segregate independently. This is the basis of Fisher's explanation for the relative advantage of sex [9]. The data also show that the average initial rate at which clonal alleles sweep is essentially the same as sexual alleles for those alleles that survive drift. This addresses another possible source of difference between sexual and clonal reproduction.

Having addressed changes in allele frequency, we superimpose contemporaneous data showing changes in adaptation score. These data are for alleles that achieve a maximum frequency that exceeds 0.1. These alleles all have their highest adaptation scores in the generation in which they come into existence. This reflects the fact that the negative frequency-dependent response has yet to emerge in response to the impending success, albeit temporary, of the emerging allele. The inevitable fall in the adaptation score is initially quite slow, but falls rapidly as the parasite response is increasing. Clonal alleles distinguish themselves from sexual ones mainly in the length of the tail that, even with three loci, is showing evidence of marked interference with selection.

### *Optimizing host mutation rates and the effects of polymorphism and locus number*

This section describes the outcome when data, used in Figure 2, on the interaction between host and parasite mutation rates, are redrawn in Figure 6 to show the response of host populations to fixed parasite mutation rates. We extend these data to include examples of the effect of locus number. As with all data generated by the simulations, a far greater number of parameter values have been tested. The four graphs in Figure 6 are representative and chosen to illustrate key points.

All four graphs in Figure 6 show that host populations optimize their fitness when their mutation rates approach those set by the parasite. There is a useful theoretical result for haploid clones [23] that indicates optimum mutation rates are the reciprocal of the interval, in generations, between changes in the environment. These should be $10^{-3}$ (a,c) and $10^{-2}$ (b,d) respectively for the data shown in Figure 6. The values coincide remarkably with those shown in the graphs obtained using a method employing stochasticity and a simple heuristic based on mutation and selection. These and other data confirm that parasites that mutate rapidly impose high rates of mutation on host defences as a condition of hosts maintaining their own fitness. Figure 6(a) shows that clonal and sexual populations progressively approach an adaptation score that is close to unity for both populations as their mutation rates rise. There is a small mutational load at these mutation rates under the conditions of the simulation, and this mutational load can be increased as selection is relaxed. The difference in adaptation scores between clonal and sexual populations at large lag loads is attributable to the Kirkpatrick and Jenkins (KJ) effect [5], covered in previous sections of these Results.

Figure 6(b) shows the effect of increasing parasite-mutation rate by 10-fold. The host now shows adaptation optima at higher host mutation rates, as expected (see previous discussion), and greater polymorphism. Since underlying attenuation of the KJ effect is likely to occur as adaptation improves, the increased gap that emerges between clonal and sexual population fitnesses is potentially dominated by the increased interference with selection in clones. The effect is to confer higher net fitness on the sexual population for most of the host-mutation range.

When the number of host loci is increased at the lower parasite mutation rate used in Figure 6 ($10^{-3}$), there is a small difference in maximum adaptation scores between clonal and sexual populations that is consistent but remains unexplained. Intermediate parts of the curves show a greater disparity in adaptation between sexual and clonal populations than in Figure 6(a), but this is attributable to a greater loss of adaptation in the clonal population caused by increasing polyclonality.

Figure 6(d) illustrates the outcome with a higher mutation rate ($10^{-2}$) and a larger number of loci (20). There is a marked collapse in the adaptation scores of the clonal population as effective allele number rises. At large loads, where maladaptation is substantial, this is likely to be due to a combination of the KJ effect and polyclonality. By contrast, the greater fitness of the sexual population at adaptation optima is likely to be due almost entirely to interference on selection in clones caused by widespread polyclonality. Sex is now fitter over the entire range of host mutation rates.

### Testing Maynard Smith

One way in which Maynard Smith argued a cost of sexual reproduction used an example in which a sexual mother was converted to a clonal mother by mutation [3]. The effect of the conversion is that the clonal mother, immediately after conversion, has the adaptation she enjoyed as a sexual mother, but now has twice the fecundity, and therefore twice the fitness. Left to her own devices, she and her descendants will rapidly outbreed the sexual population and drive it to extinction. We have shown in simulations of separate sexual and clonal populations that a clonal population can have a much greater load than the sexual under the same conditions and this disparity provides a mechanism by which sex can, in principle, prevent invasion by clones. In the model, the clone is chosen from the pool of best-adapted sexual mothers. This gives the clone an adaptation score that, at steady state, is better than a clonal mother would normally enjoy. The outcome of inserting a clone is an empirical matter, however, and rests on the size of the difference in adaptation and the rate at which the parasite environment causes loss of adaptation. Too small a difference in adaptation (that is, too small a load) and sex cannot sustain a fitness advantage in the face of a clone; too slow a response from the parasite and the clonal population expands rapidly at little cost. Figure 7 shows data for three populations ($N = 10^2, 10^3, 10^4$), where the numerical value of $\mu_p / \mu_h$ (= 1.0) is sufficient to give sex a fitness advantage at steady state. Clones are generated by conversion and the simulation tracked to completion. At least two predictions follow from this brief outline. First, the parasite mutation rate, $\mu_p$, needs to be above some threshold that causes rapid deterioration in the environment, and second, the outcome should show some sensitivity to population size, since larger populations provide for a larger number of generations to extinguish the descendants of a clone. Because the sexual female chosen for conversion to a clone is the best adapted, it inevitably means she is rare and, initially, without a negative frequency-dependent response from the parasite. Larger host populations provide a higher number of generations within which to generate an adequate negative frequency-dependent response. Both these predictions are borne out by the data in Figure 7.

**DISCUSSION**

The challenge for any theory of obligate bi-parental sex is to show that there are biologically relevant conditions under which individual selection in diploid sexual populations may lead to higher population fitness than in clonal ones and, further, to demonstrate that sex can hold its own or beat cloning in mixed sexual and clonal populations using fitness-dependent selection of individuals for reproduction.

We chose to pursue the idea that sex might have an advantage in dynamic environments that changed sufficiently to create differential effects on mutation and selection between clonal and sexual populations. More specifically, we were interested in environments, such as those produced by parasites, where there are ongoing adverse effects due to negative frequency-dependent selection on hosts. Hamilton and colleagues were the first to put forward the idea that sex is an adaptation to resist parasites [16,17], and they developed a simple heuristic for recombination and selection in haploids. It is unclear that this model overcomes the two-fold cost of males. We felt it would be of interest to test the idea in a mutation and selection model in diploids.

Mutation-selection models have proven challenging to analyse [18], particularly where the environment is changing (see the extensive treatment in [20], Chap. 4). Environmental change can take several forms. It can take the form of a random walk that intermittently imposes directional selection but, equally intermittently, relaxes selection. There is fluctuating selection, which is semantically ambiguous. It frequently means oscillatory movement, possibly erratic, between two states (+/-) but could, at a stretch, be extended to random walking. And there is negative frequency-dependent selection, which is capable of imposing ongoing selection pressure in a large allele landscape, even if the precise direction of selection shifts stochastically. It is not clear that any theoretical treatment, to date, treats the consequences of negative frequency-dependent selection on a diploid population. Rather than struggle with what appears to be a daunting analytical challenge, we chose instead to explore the possibility that a simple heuristic approach could provide an adequate framework to study sex in the context of mutation and selection.

Any heuristic method is *sui generis*, and relevance stems from its ability to provide some insight that is potentially unavailable by other means. The heuristic method we use is extremely simple, and open to refinement using what, in AI terms, are evolutionary algorithms. The current heuristic method captures a number of expected features of mutation and selection in natural populations. These include: (i) recapitulation of the rate effect first identified by Kirkpatrick and Jenkins [5]; (ii) an optimum mutation rate for clones in a dynamic environment that matches theory; (iii) expected birth and death of alleles in sexual and clonal reproduction, together with plausible qualitative behaviour of individual alleles; (iv) qualitatively plausible behaviour for allele distributions, moving from dominance of stabilizing selection to effects of directional selection as environmental mutation increases; (v) predicted qualitative behaviour of increasing mutation rates and locus number in hosts; (vi) predicted qualitative effects of injecting fit clones into sexual populations; and (vii) emergence of conditions under which sex and cloning can plausibly be in fitness balance. In addition to these features, one could add the behaviour of newly formed alleles under conditions of drift, and their occasional escape from drift into the body of the population where they come under selection.

If we accept, for the purpose of argument, that this heuristic approach has some utility, we can use it to explore behaviour of populations under negative frequency-dependent selection. What emerge are two sets of conditions that support sex. One is an environment of substantial lag load. This load can be generated at a wide range of absolute mutation rates, provided the diploid retains a substantially sub-optimal mutation rate. The advantage to sex is due to the effect identified by Kirkpatrick and Jenkins for a single locus [5]. However, the advantage disappears when the lag load disappears and, for low mutation rates, this is where the host mutation rate approaches an optimum.

The second environment that supports sex is one that emerges conjunctively as the number of loci under independent selection increases and host mutation rates evolve to provide resistance to parasites. Significant mutational demands on hosts arise because parasites can mutate rapidly and tolerate high mutational loads. One consequence of higher mutation rates is the potential for extensive host polymorphism if clearance of deteriorating alleles is not efficient.

The size of the advantage conferred by a mutation is important. At one end of the spectrum, we have mutations of major effect that dominate the fitness of the clone they are adopted by. The clone displaces its rivals because of its higher fitness and the population becomes monoclonal. There is no wastage of advantageous mutations if sweeps are complete, and cloning suffers no disadvantage relative to sex. At the other end of the spectrum, the advantageous mutation is of very small effect and the outcome is determined stochastically by drift between the various clonal candidates. Many advantageous mutations are wasted, but only if the clonal population carries an effective allele number greater than one. It is not easy to generate a high effective allele number in a static environment because the population remains largely monomorphic (Figure 4 (b)); less fit outliers, generated by mutation, are constantly removed by stabilizing selection. It is only when the environment itself mutates using negative-frequency dependent selection that inefficiency in truncation of deteriorating alleles emerges. The factors that exacerbate these inefficiencies are faster rates of mutation, and larger numbers of host loci.

What the heuristic described here suggests is that, subject to an adequate mutation rate and number of loci engaged, a gap opens up between the fitness of sexual and clonal populations that is sufficient to confer a fitness advantage to sex at optimum host mutation rates. A particular type of mutating environment is essential. In a fixed environment, stabilizing selection ensures that sex and clones have effectively the same adaptation scores, even when the number of loci is large, and clones are much fitter. Equally, environments that behave as random walks are ineffective at producing an advantage to sex, because they keep relaxing selection pressure on clones, allowing them to eliminate their larger lag loads and catch up with sex.

By contrast, the adaptation gap that opens up with negative frequency-dependent selection yields greater inefficiency in clonal clearance, numerous clonal lines emerge that are reproductively isolated. These potentially waste advantageous mutations. Sexual populations, which form a single reproductive population, avoid wastage because they can shift advantageous mutations efficiently to fitter backgrounds through independent assortment and are only marginally affected. This is the essence of Fisher's explanation for sex, outlined on p.123 of his book [9], although, as pointed out in the Introduction to this paper, his is only a verbal explanation and the two-fold cost of males appears to be unrecognized. Fisher's explanation requires interfering polymorphism in clones, and this is provided through a Red Queen driving higher wastage of advantageous mutations in clones and slower selection of deteriorating alleles. Hence the title of the paper.

Does the KJ effect still operate at high mutation rates when host populations have optimum mutation rates? One cannot be certain. There are obvious loads with high mutation rates and large numbers of loci but these are less obviously lag loads in the original sense [6]. Rather, they are loads created by declines in the efficient removal of obsolescent alleles that are very marked in clones. Opportunities for adoption of new advantageous mutations become restricted, and the KJ effect much weaker as a result. One prediction that has yet to be tested is that, in populations with large numbers of loci, the actual and effective number of alleles rise disproportionately in clones relative to sex, because of inefficient clearance, and the rate of adoption of new mutations falls.

Since mutation rates in the genome overall are much too low to accommodate high mutation rates at defence loci, the heuristic suggests that there has to be a two-tier genome at the mutational level. This is an idea canvassed in [24] with respect to bacteria, but which could be extended to adaptive immunity in mammals, for example.

At this point, we need to ask whether either of the conditions identified using the current heuristic approach are remotely plausible. On the issue of lag loads, we simply do not know whether the size of lag load required is consistently attainable. The most likely cause of sustained lag loads is not a sporadically and slowly changing abiotic environment, but rather a biotic environment whose rate of change exposes higher organisms, such as eukaryotes, to the demanding task of evolving optimum mutation rates that are high. This second condition overlaps with the first, but extends to instances where the lag load difference between sex and clones occurs at optimum, or near optimum, mutation rates.

Here we turn to the biological data.

*Levels of selection.* Our own species, *Homo sapiens*, presents a demanding test of the ideas presented in this paper, because of its low fecundity and long maturation time to reproductive adulthood. Parasites reproduce some $10^3 – 10^4$ times faster than humans when generation times are compared, and we have emphasized the need for host defences to mutate at rates commensurate with those of parasites. To this end, mammals, including ourselves, have evolved a rapidly-adaptive immune system that has many of the properties of viral pathogens: it is targeted, it has high mutation rates, and can tolerate high mutational loads. Other parts of the adaptive immune system, notably the HLA system, sit under this first line of defence. Remarkably, the HLA system is sufficiently protected that, despite its germ-line encoding and central position in the adaptive immune response, it provides an antigen presentation system that resists a considerable amount of parasite evolution during an individual lifetime. Nevertheless, the HLA system shows the highest rates of mutation of any mammalian locus, particularly the MHC Class I system that presents viral antigens, suggesting considerable selective pressure and a need to generate novel molecular variants. As we discuss shortly, disease was the major cause of mortality in our evolutionary forbears until improvements in public sanitation occurred in the 19[th] century and levels of mortality were surprisingly high.

*Pathogen evolution.* Extensive temporal data on pathogen evolution and its timelines are still sparse for the most part, but there are important data for the influenza viruses [25-28]. Here, for example, the data for H3N2 show global sweeps approximately every 6 months [28]. The rate of mutation is much higher, but there are many evolutionary dead-ends. Full sweeps represent the emergence of major escape mutants. There are several mechanisms by which these mutations arise but these lie beyond the scope of this discussion. Importantly, there is abundant evidence from the size and complexity of phylogenetic trees to suggest that the influenza molecules under selection (i.e. the haemagglutinin and the neuraminidase) occupy extensive allele space. An important mechanism driving virus evolution are host defences that drive the emergence of pathogen escape mutants which are novel, for which yet further evolution of host defences is required. This has important consequences for any simulation because it justifies the use of extensive allele space for the parasite and, given the inverse matching allele (IMA) system used in the simulation, to the host.

There are, perhaps, 10 – 20 major infectious diseases that were the scourge of our immediate ancestors. Although influenza viruses represent some of the fastest evolving pathogens and may be atypical, it is not inconceivable that major new infectious challenges to humans emerged historically on a monthly basis, providing considerable selective pressure.

*Selection.* The simulation we describe relies on an environment that is not only characteristic of that inflicted by parasites but where parasites, in the evolutionary past, are a major source of morbidity and mortality. Reliable data on mortality of earlier human populations is slowly emerging. Swedish data on human populations are still some of the best on record, despite the need for care in using them. Complete population data for all individuals is available for each year from 1751 [29]. Mortality in prehistoric hunter-gatherer populations was slightly worse than the Swedish populations [30]. If we assume that menarche in females began at 20 and the menopause occurred at 50, then 44% of an original cohort of breeding hunter-gatherer females died before their reproduction could begin and a further 24% died during their breeding years. If sterility is taken as 16% of the population (the modern figure), then over 60% of primitive human populations had zero or limited reproductive fitness. Epidemics resulted in even greater mortality. If we assume that most deaths were due to disease, the truncation of human populations applied severe selective pressure on host and defences. Other species are much more fecund than humans and can withstand harder selection.

*Polymorphism.* A consequence of a rapidly evolving parasite environment is that host defences also have to evolve at a rate that offers protection. A major point of difference between our model and those of others is in assuming very large evolutionary landscapes for pathogens, and large allele landscapes for host defences that can only be accessed fully by mutation, notwithstanding the mutational load this generates. In these circumstances, failure to offer adequate beneficial mutations invites reduction in host adaptation as parasites increasingly circumvent host defences and increase the host population's morbidity and mortality. Hosts have no direct control over the mutation rate of parasites, which could, in principle, evolve to high levels, not least because they can potentially sustain high mutational loads. Increasing mutation rates increases the amount of polymorphism in a population by shifting mutation/selection balance. Polymorphism therefore becomes a potential signature of a response to a rapidly changing, hostile environment, even in sexual populations, because of the increasing inefficiency of allele clearance.

One of the clearest examples of extensive polymorphism in higher eukaryotes is the MHC locus in vertebrates [31,32], whose function is to present pathogenic peptides to the adaptive immune system. Pathogens rapidly evolve escape mutants to avoid presentation to host defences, placing the HLA system, in particular, under severe selective pressure, particularly the HLA-A, B, and C loci that present viral peptides. Although figures from the European Bioinformatics Institute record the most polymorphic HLA loci as having over 1000 alleles each [31], recent estimates for the number of alleles at HLA-A, HLA-B, HLA-C and HLA-DRB1 loci currently exceeds 3 million [33], and the number of theoretically possible alleles is two orders of magnitude higher. Prospective allele space is therefore very large, even though many of the current alleles are rare, likely reflecting the very large recent increase in global human population. The HLA system is germ-line encoded and has to provide utility over the lifetime of the individual. It is partly protected by the antibody system, which has many features in common with parasites: it evolves rapidly, it mutates extensively using cassette assembly and somatic hypermutation, and it can withstand high mutational loads. This may actually drive the adoption of sex in some parasites [34].

Another system that potentially involves polymorphism in host defence genes is that between *Arabidopsis thaliana* and Downy Mildew (*Hyaloperonospora arabidopsidis (Hpa*)), where pathogen-derived effector molecules form a class of diverse and fast-evolving proteins. Recent evidence [35,36] suggests that a highly adaptive pool of defence genes in the host (*R* genes) of *A. thaliana* has the potential both to maintain recognition of parasite effector molecules as well as rapidly establish new recognition specificities. Other examples of extensive polymorphism at defence loci could be cited.

*Mutation rate.* Our heuristic rests, in part, on the relative benefit to sex of adoption of advantageous mutations, and the advantage is strengthened at high systemic mutation rates that increase the level of polymorphism, because it causes interference with selection in clones. However, deleterious mutations normally outweigh beneficial mutations and this could create serious difficulties for the model in low fecundity species, such as scarabs and *Homo sapiens*, that may not readily support the mutational load. This objection would be blunted if mutations were of conditional benefit whose advantage rested on contemporaneous conditions; that is, a benefit whose magnitude rested on conditional fitness. This is, in fact, the basis upon which the current simulation is built. There are no alleles in our allele space that cannot confer an advantage under some circumstances. The least fit allele in our system is simply one that it is temporarily unable to provide an effective defence. A library of conditionally fit alleles that can be tapped at random still carries a cost, but this is potentially outweighed by the benefit that accrues from a higher probability of adopting an advantageous mutation.

There is now increasing evidence that MHC/HLA diversity, as outlined above, may be predominantly an outcome of gene conversion using cassettes that transfer existing sequences of adaptive value [32,37,38]. However, a study of the San Nicolas Island fox (*Urocyon littoralis dickeyi*) suggests that mutation can remain a potentially important mechanism [39]. This fox population has recently passed through an extreme population bottleneck of less than 10 individuals within the past 10-20 generations, creating a genetically monomorphic population over the bulk of the genome. By contrast, the MHC locus has rapidly regenerated considerable polymorphism by mutation. From this, we can conclude that high rates of mutation are possible at specific sites in an otherwise quiescent genome, a scenario compatible with our simulation. Moreover, the selection coefficients on MHC loci are greater than 0.5 [39], similar in magnitude to those operating in human hunter-gatherer populations. Another recent example of rapid adaptive evolution of the MHC locus is found in [40] and rapid adaptation through adoption of advantageous mutations is described in [41].

*Experimental evidence.* A number of experiments supporting Red Queen models have been published over the past 25 years [42-48] and a recent review of these models [49] contains additional references to experimental data. Studies of the New Zealand snail [43], *Potamopyrgus antipodarum*, are of interest, within the context of the current simulation, for at least two reasons that may be interlinked: (i) the timeline for their coevolution with a sterilizing parasite species of *Microphallus* is possibly slow compared with viral and bacterial pathogens, and (ii) mixed populations of sexual and asexual individuals are common in their natural habitat and appear broadly stable. As we have shown in our simulation, there are parasite mutation rates at which sexual and clonal populations have the same fitness and mutual co-existence should be stable. An increase in parasite mutation rate in this environment would then favour sex, and a decrease would favour clonality. Work on *P. antipodarum* had previously suggested that asexual populations prevail in areas associated with low risks of infection, as expected (see [49] for references). Two other studies [45,46] have explicitly identified the adoption of advantageous mutations as the basis for adaptation. Finally, a recent paper [48] suggests that sex enhances adaptation by allowing independent selection on advantageous and deleterious mutations. This is a major function of sex in the current simulation, with clones suffering adversely when this independent selection is made less efficient.

*Final comments.* The heuristic we use identifies two mechanisms for the maintenance of sex. Both differ from current mechanisms, although both re-instate ideas that have previously been circulated and discarded. Although the two mechanisms add to an enormous pile of alternatives, they do make some fairly clear predictions. Indeed, we would argue that some evidence is already consistent with the model we propose, and further evidence probably exists in the literature and is capable of assembly. Particularly gratifying in the heuristic is the identification of conditions under which sex and cloning can co-exist, because this type of co-existence, or easy transition between sex and cloning, is widespread. It forms the basis of the balance argument put forward by Williams.

Returning to our original point of departure, namely, the two-fold cost of males identified by Maynard Smith [3], we can see it is only one term in a cost equation. It lacks compensation in static and slowly evolving environments, where clones win, but in dynamic environments that are persistently antagonistic and impose lag loads, the costs of males and outcrossing in sexual reproduction are less than those inflicted by the environment on clonal fitness. The position is made worse for clones as the number of loci increases, and various forms of selective interference are recruited. The comparatively higher fitness of sexual progeny reflects faster adaptation to environmental change, but not for the reasons normally cited. As a result, sex is an Evolutionarily Stable Strategy (ESS) in environments characterized by Red Queens, particularly Red Queens that run fast, but is not an ESS in equilibrium or near-equilibrium environments.

**METHODS**

We use a heuristic approach that employs a genetic algorithm within the class of artificial intelligence (AI) methods known as evolutionary algorithms. Implementation of the genetic algorithm adopts a Wright-Fisher approach using random mating and discrete non-overlapping generations in a single population without geographic effects.

Steps within the implementation are:

(a)     establish a finite initial population of individuals;

(b)     evaluate the adaptation or fitness of each individual in that population;

(c)     apply operators (power function, adaptation/fitness threshold, etc.) where applicable

(d)     select best-adapted or fittest individuals stochastically for reproduction (i.e. choose parent(s)), using an adaptation- or fitness-dependent metric

(e)    generate haploid gametes for sexual reproduction

(f)    breed next-generation individuals by syngamy (sexual reproduction) or mitosis (cloning)

(g)    apply operators (mutation, crossing-over, etc.) where applicable

(h)    repeat steps (b) – (g) until termination (time limit, adaptation steady states, etc.)

These steps differ slightly from those in a conventional AI evolutionary algorithm, where it is a candidate solution to an optimization problem that plays the role of an individual in competition with other optimization solutions.

The aim for the heuristic method is analysis of the behaviour of a host diploid population under attack from a haploid parasite. The parasite attack is driven by negative frequency-dependent selection on the host, which responds by developing effective resistance alleles within a defence locus. Alleles in host and parasite are conventional number strings (16 x 1-bit), which give 65,536 possible alleles. The diploid host locus encodes symbolically a defence protein with 32 x 1-bit sites (= 2 x 16-bit alleles) for binary (0,1) mutation in either strand. The haploid parasite locus encodes symbolically an essential protein that enables the parasite to evade host surveillance, progressively increasing host allele morbidity and mortality. The adaptation of each allele (step (b) above) is a numerical score obtained by inverse matching with reference to opposing alleles in the other species. For example, a host allele number string, `0110010111011000,` has a match score of 10 when measured against a parasite allele number string, `0011000110111010`. The parasite allele has the complementary score of 16 – 10 = 6. Thus, adaptation of the alleles of one species increases as adaptation of alleles in the other declines, and *vice versa*. This is intended to reflect, on a simple numerical scale, the ability of the host to defend itself and the ability of the parasite to evade host defences. Individual allele adaptation scores are calculated against the weighted population of alleles in the counter-species. Adaptation scores for individual diploid loci are the arithmetic mean of the allele scores in these simulations; there is no heterozygote advantage.

Adaptation scores are calculated for each allele, parasite and host, for each generation, including average population score. Populations are fixed in size for each simulation and remain unchanged during each run. Individuals are chosen stochastically for mating (step (d) above) using a function of their adaptation score, either the score itself or a power function of the score, to determine the probability of selection for reproduction. The exponent for the power function is either 1 or 5 for the simulations described here. We have not used adaptation or fitness thresholds in the current simulations.

For sexual reproduction (step (e) above), which is restricted to host diploids, a diploid parent is chosen randomly on the basis of adaptation score and one allele is chosen at random for transfer to the next generation. This constitutes formation of a haploid gamete (sperm or egg). A second parent is selected on the same basis and, again, an allele is chosen at random. Parents are retained in their own generation, where they can engage in reproduction on a second or subsequent occasion. Parents cannot breed with themselves. Gender of offspring can be assigned at random to create two sexual populations, but it makes no difference to the computation since the two gender pools are essentially the same in their composition. For clonal reproduction, a selected parent is copied by mitosis. We make a clear distinction between adaptation, which reflects fitness for purpose in both host and parasite, and fitness, which is a multiplicative function of adaptation and fecundity. For clones, two copies are inserted in the next generation for every one supplied by sex. This imposes a two-fold fecundity disadvantage on sexual reproduction, where only 50% of the progeny are female.

Mutations are normally point mutations occurring with rate, $\mu$, that represents the probability that an allele will undergo a 1-bit alteration in an allele per generation. It is possible to arrest evolution in the host population while the parasite population undergoes a number of reproductive cycles. This progressively optimizes the adaptation score of the parasite through iterative scoring, exposing the host to an abrupt step-jump increase in parasite virulence when the host is reactivated. The host has effectively suffered imposition of a step lag load that may be substantial. The simulation can also produce random walks in the parasite, which has the effect of imposing intermittent lag loads on the host, as opposed to the continuous lag loads imposed by negative frequency-dependent selection.

Diploid genotypes can also contain multiple loci. Each locus has its own parasite that acts to optimize its own adaptation score and minimize that of the host at that locus. Adaptation scores for each locus are established as for single loci, but adaptation scores for the overall genotype are established multiplicatively. For a fully sexual individual, each locus is uncoupled from all others, to reflect independent assortment of each locus. The simulation contains a toggle, however, that enables replacement of independent assortment at each locus with a haplotype in which an allele from each locus is fully coupled with alleles at other loci and remain so for the duration of a run. Haplotypes undergo a birth-and-death cycle in the same way as alleles at single loci. Data available from each run include effective allele number, allele frequency, allele identity, and allele adaptation scores, together with derived data.

Simulations of steady states are run for extended periods of time, the lengths of which are determined by the time taken to reach and assay the steady state. Populations with high mutation rates respond rapidly. Longer burn-in times and sampling periods used for slowly mutating populations are usually extended *pro rata* from faster simulations, with a ten-fold decrease in mutation rate resulting in a 10-fold increase in the length of the simulation. The longest simulations have used burn-in periods of >20 million generations and sampling periods of >100 million generations. The ability of sexual diploid populations to resist invasion by clones was tested by injection of clones after a suitable burn-in period. In the current simulation, a sexual female of the highest adaptation score was converted to clonality. Her fitness doubles because she is twice as fecund as her former incarnation as a sexual mother. Simulations were repeated, typically for 500-1000 runs.

The software is written in C and is available from the authors.

**ACKNOWLEDGEMENTS**

This research was supported by New Economy Research Fund grant UOOX0809 from the New Zealand Foundation for Research, Science and Technology (FoRST), now administered by the New Zealand Ministry for Business, Innovation and Enterprise (MBIE). The funders had no role in study design, data collection and analysis, decision to publish, or preparation of the manuscript.

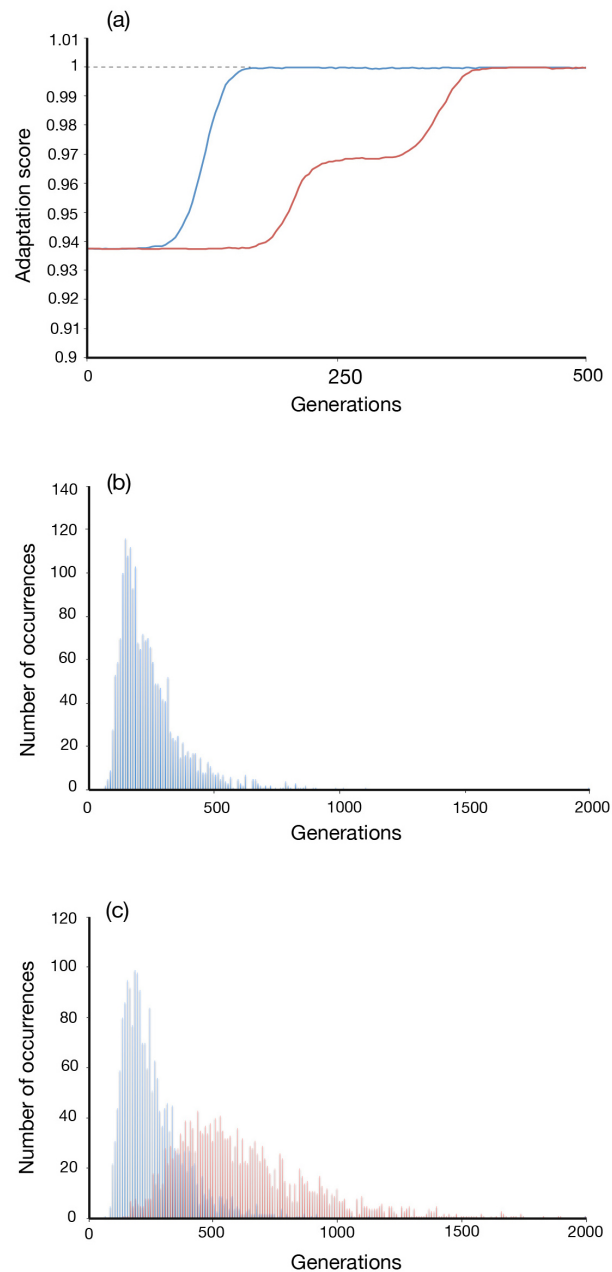

**Figure 1.** Computer simulations of adoption of advantageous mutations.

(a) Computer simulation showing adoption of advantageous mutations at a single diploid locus. The starting populations are sexual (blue) and clonal (red) (population size, $N$, = 1500; host mutation rate, $\mu_h$, = $10^{-4}$ bits/allele/generation). Each starting population is uniformly homozygotic and maladapted by one bit per allele (starting adaptation of 0.9375). The parasite environment is uniform and fixed for the duration of the simulation (that is, mutation rate, $\mu_p$, = 0). The sexual population shows a single sweep to full adaptation following adoption of an advantageous mutation, whereas the clonal population proceeds through two sweeps, reflecting the need for adoption of two mutations to proceed to full homozygous matching of the parasite. The target size for the second mutation in the clonal heterozygote is half that for the first mutation, halving the rate of adoption. This is the basic result shown by Kirkpatrick and Jenkins [5]. (b) Distribution of times to 25% adoption of advantageous mutation for the sexual population in (a). Average time to 25% adoption, in this simulation = ~231 generations. (c) Distribution of times to adoption of advantageous mutations for the clonal population in (a). Average time to 50% adoption of first mutation, in this simulation = ~250 generations for first mutation, and ~681 generations for second mutation. The longer average time for adoption of the second mutation reflects the halving of the mutation target size.

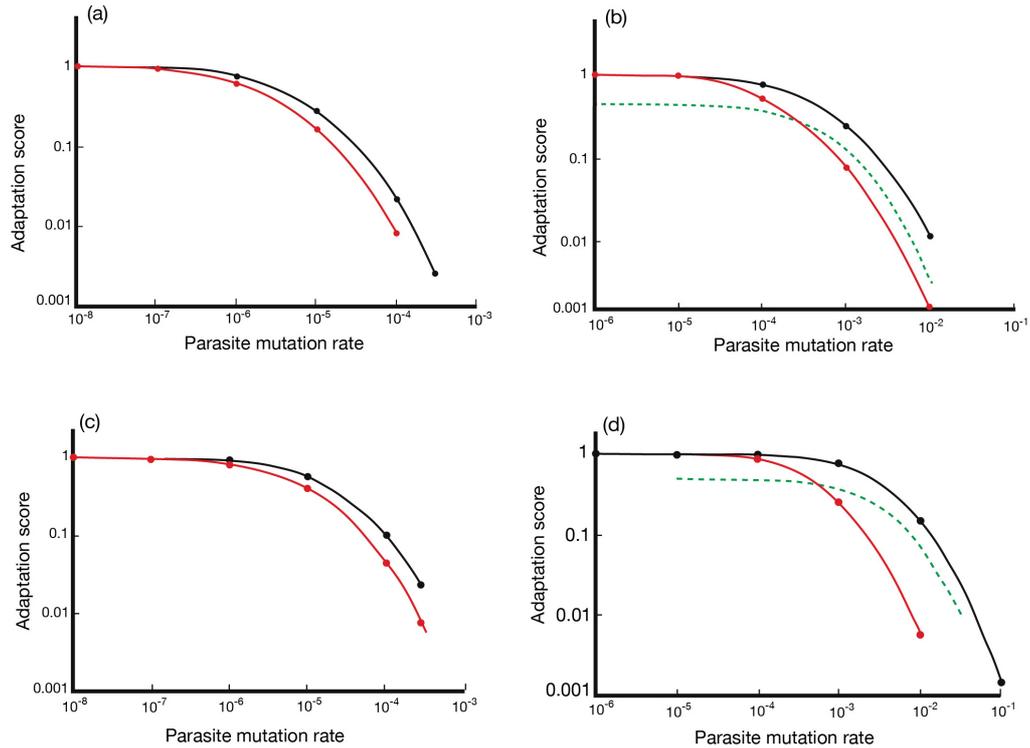

**Figure 2**. Adaptation profiles for sexual and clonal host populations at two host-mutation rates.

(a) A sexual population ($N$ = 100; $\mu_h$ = $10^{-6}$ bits/allele/generation; $\mu_p$ = $10^{-8}$-$10^{-3}$ bits/allele/ generation; number of loci, $L$ = 10) (black curve), and a clonal population under the same steady-state conditions (red curve). The simulation consistently gives higher population-average adaptation scores to the sexual population when compared with the clonal population. The value of $\mu_p/\mu_h$ is progressively reduced in each curve by increasing $\mu_p$ at constant $\mu_h$. This changes the opportunities for adoption by target populations of advantageous mutations that offset the antagonistic evolution of the parasite. This figure shows that the rate advantage accruing to sex in the adoption of advantageous mutations (shown in Figure 1) persists in temporally complex environments when numerous allelic species are in play and there is potentially more than one mutation occurring in each generation.

(b) A sexual population ($N$ = 100; $\mu_h$ = $10^{-3}$ bits/allele/generation; $\mu_p$ = $10^{-6}$-$10^{-2}$ bits/allele/ generation; number of loci, $L$ = 10) (black curve) and a clonal population under the same steady-state conditions (red curve). The differences in population-average adaptation scores between sexual and clonal populations for $\mu_h$ = $10^{-3}$ are greater than for $\mu_h$ = $10^{-5}$ (shown in Figure 2(a)). This is attributable to raised polymorphism at higher mutation rates, which disproportionately affects the clonal population. The green curve represents the adaptation score for the sexual population adjusted by a factor of 0.5 to reflect its lower fitness relative to the clonal population due to the 2-fold cost of males. The green curve and the clonal adaptation curve can now be used to indicate relative fitness. The crossing-point marks the parasite mutation rate, $\mu_p$, at which sexual and clonal populations have equal fitness. When the parasite population mutates more slowly than this rate, the fitness of the clonal population is higher; when the parasite mutates more rapidly, sex has the higher relative fitness. (c) A sexual population ($N$ = 3000; $\mu$ = $10^{-6}$ bits/allele/generation; $\mu_p$ = $10^{-8}$-$10^{-3}$ bits/allele/ generation; number of loci, $L$ = 10) (black curve) and a clonal population under the same steady-state conditions (red curve). The data are similar to those in Figure 2(a) but the larger population has produced a slightly larger difference in steady-state adaptation scores for sexual and clonal populations. (d) A sexual population ($N$ = 3000; $\mu$ = $10^{-3}$ bits/allele/generation; ; $\mu_p$ = $10^{-6}$-$10^{-1}$ bits/allele/ generation; number of loci, $L$ = 10) (black curve) and a clonal population under the same steady-state conditions, (red curve). The data are similar to those in Figure 2(b) but the larger population has produced a greater difference in steady-state adaptation scores for sexual and clonal populations. The green curve has the same status as in (b). The clonal population is also subject to interference with selection, due to increased polymorphism at the high mutation rate. This interference serves further to reduce the adaptation score.

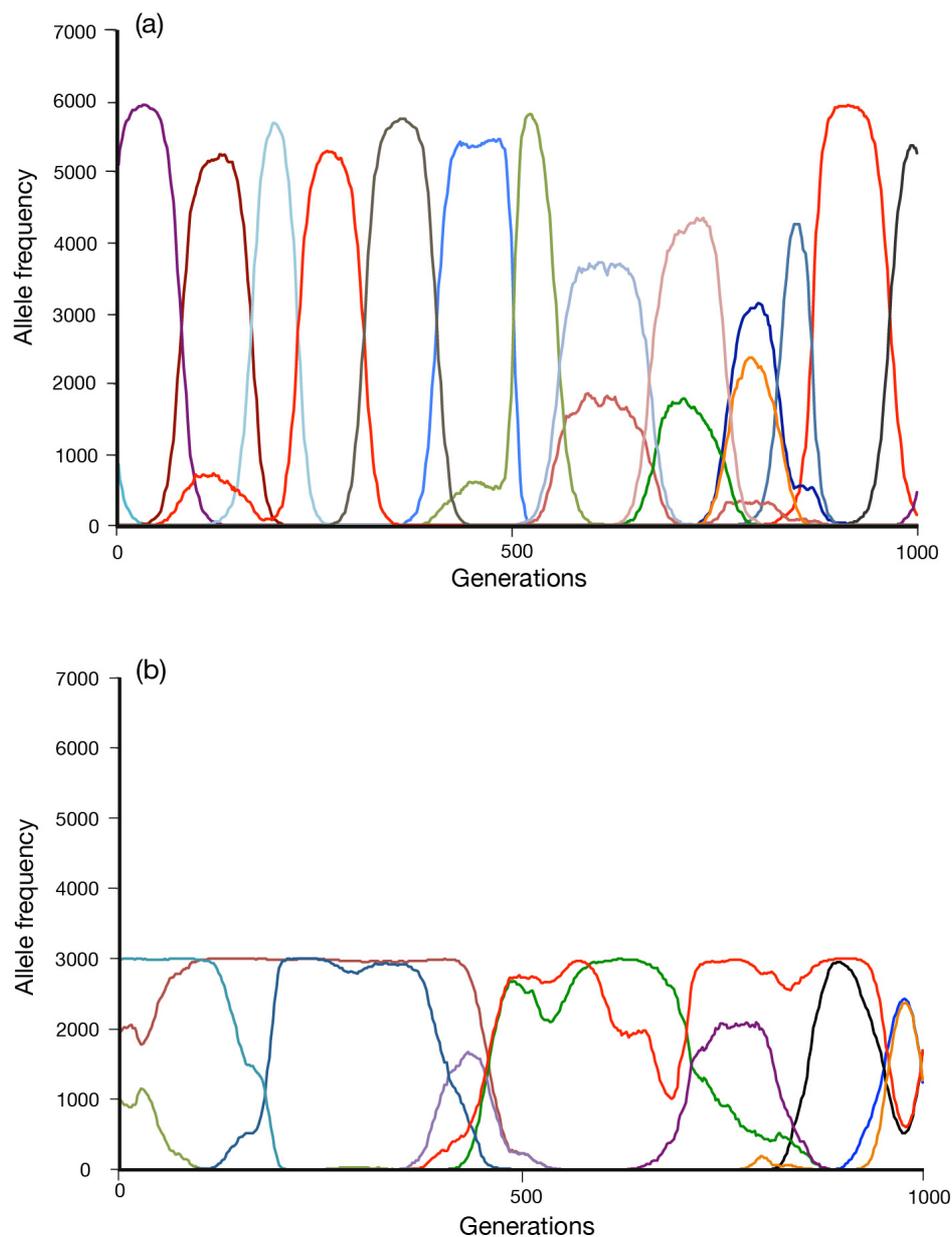

**Figure 3.** Allele behaviour of sexual and clonal populations under lag load.

(a) A sexual population ($N$ = 3000; host mutation rate, $\mu_h$ = $10^{-4}$ bits/allele/generation; parasite mutation rate, $\mu_p$ = $10^{-2}$ bits/allele/generation; number of loci, $L$ = 3) at steady state. The population adaptation score for each locus is an average of 0.726, and the overall population adaptation score is 0.382. Allele frequencies shown are for those alleles whose maxima exceed a frequency of 600. At these rates of host and parasite mutation, alleles have a typical lifetime of ~100 generations and the majority of sweeps reach almost complete homozygosity. There is also an occasional incidence of limited balancing polymorphism. Different colours represent different allele species. Most alleles do not cycle more than once but there are occasional instances of bimodality, indicating a temporary localization of the parasite in allele space. (b) Clonal population ($N$ = 3000; host mutation rate, $\mu_h$ = 10$^{-4}$ bits/allele/generation; parasite mutation rate, $\mu_p$ = 10$^{-2}$ bits/allele/generation; number of loci, $L$ = 3) at steady state. The population adaptation score for each locus is an average of 0.564, and the overall population adaptation score is 0.181. Allele frequencies are for those alleles whose maxima exceed a frequency of 600. Maximum allele frequencies cap themselves at 3000 under substantial lag load for reasons discussed in the body of the paper, making the clonal population fully heterozygotic under these conditions. High-frequency alleles, capped at 3000, show much longer lifetimes than their sexual counterparts, due to interference with selection. An example of this interference is the fluctuation in allele frequency associated with very long-lived alleles (illustrated by the red line). Where partial dips in frequency followed by recovery indicate early negative selection on the genotype in question, followed by an advantageous mutation at one of the other coupled loci. This brings the genotype as a whole under positive selection, reversing the impending early loss of the allele in question. Final elimination of this allele runs beyond the end of the simulation.

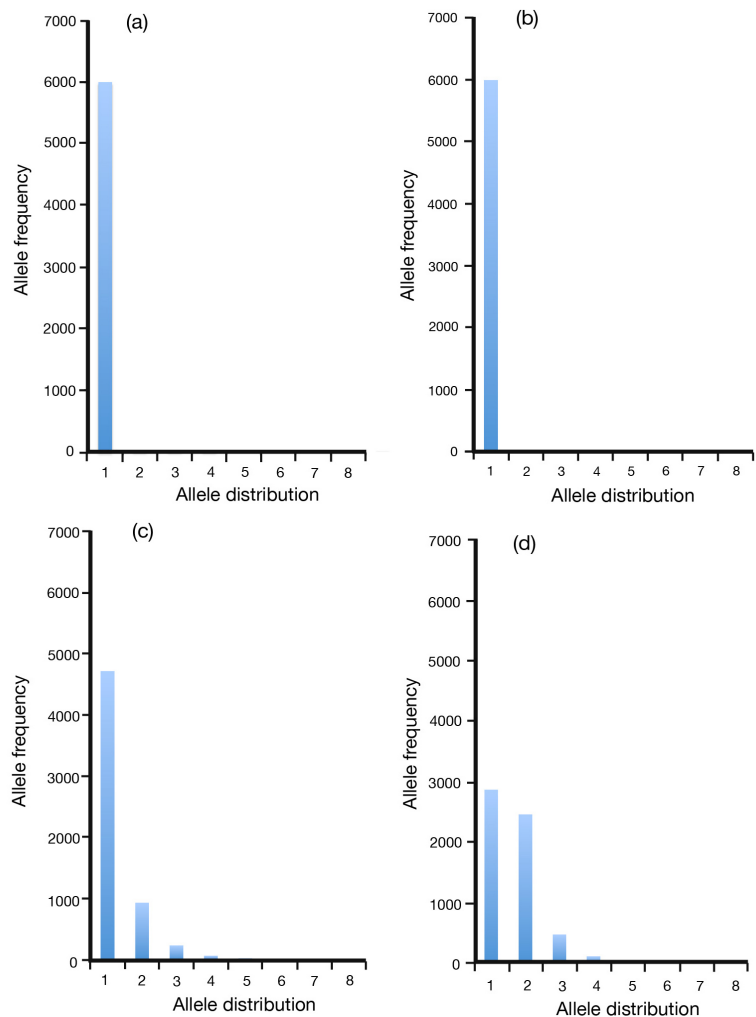

**Figure 4.** Allele frequency distributions with and without lag load in sexual and clonal populations.

(a) A sexual population ($N$ = 3000; host mutation rate, $\mu_h$ = $10^{-4}$ bits/allele/generation; parasite mutation rate, $\mu_p$ = 0; number of loci, $L$ = 3) at steady state. The population is essentially fully adapted, and monomorphic over the short and long term. There is a small standing variation of low frequency alleles that does not register on the graph. (b) A clonal population ($N$ = 3000; host mutation rate, $\mu_h$ = $10^{-4}$ bits/allele/generation; parasite mutation rate, $\mu_p$ = 0; number of loci, $L$ = 3) at steady state. The population is essentially fully adapted, and monomorphic over the short and long term. Standing variation is almost identical to that in Figure 4(a). (c) A sexual population ($N$ = 3000; host mutation rate, $\mu_h$ = $10^{-4}$ bits/allele/generation; parasite mutation rate, $\mu_p$ = $10^{-2}$ bits/allele/generation; number of loci, $L$ = 3) at steady state. Rank order frequency distribution (frequency average of commonest allele, second commonest, etc.) reflects the long run averages from Figure 3(a). Substantial lag loads increase the frequency of some lower order alleles. (d) A clonal population ($N$ = 3000; host mutation rate, $\mu_h$ = $10^{-4}$ bits/allele/generation; parasite mutation rate, $\mu_p$ = $10^{-2}$ bits/allele/generation; number of loci, $L$ = 3) at steady state. Rank-order frequency distribution of alleles reflects the long run averages from Figure 3(b). Substantial lag loads increase the rank order frequency of previously rare alleles and cap the commonest allele at slightly less than 0.5. The level of heterozygosity increases markedly when compared with the sexual population under the same circumstances, potentially trapping advantageous mutations disproportionately.

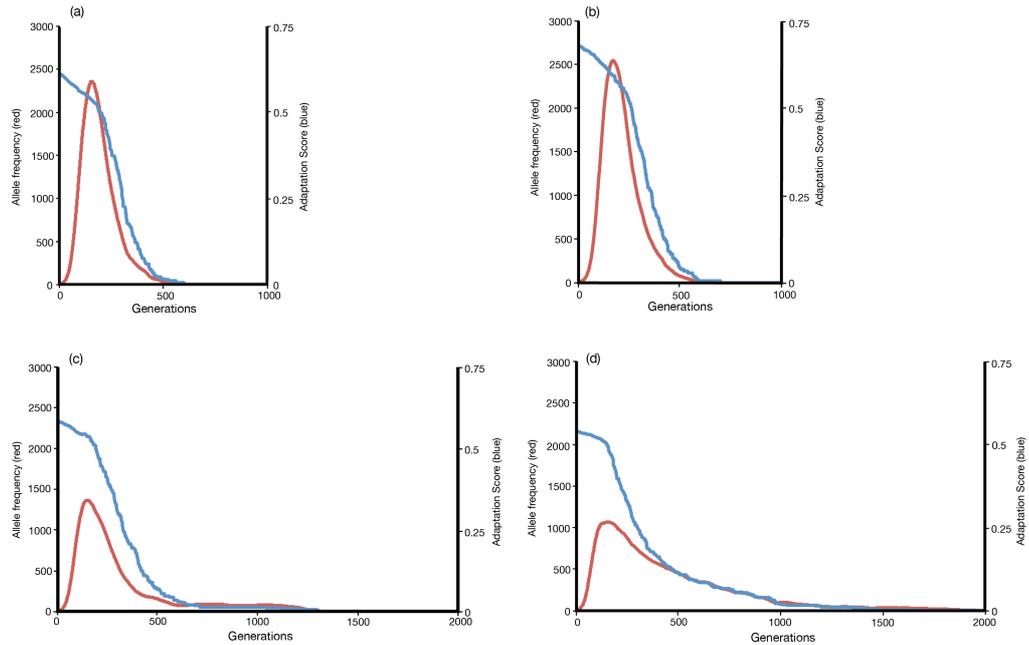

**Figure 5.** Allele frequencies and adaptation scores as a function of time.

(a) A sexual population ($N$ = 3000; host mutation rate, $\mu_h$ = 10$^{-4}$ bits/allele/generation; parasite mutation rate, $\mu_p$ = 10$^{-2}$ bits/allele/generation; number of loci, $L$ = 1) at steady state. Allele frequency reflects composite time course for 500 adopted mutations with a frequency threshold of 600 (= 10% of maximum possible frequency). Curves are synchronized for first appearance of the advantageous mutation. Adaptation is also scored from point of adoption of advantageous mutation, when scores are highest. A negative frequency-dependent response begins immediately, but loss of adaptation picks up sharply after peak allele frequency has been attained.

(b) A sexual population ($N$ = 3000; host mutation rate, $\mu_h$ = 10$^{-4}$ bits/allele/generation; parasite mutation rate, $\mu_p$ = 10$^{-2}$ bits/allele/generation; number of loci, $L$ = 3) at steady state. Similar to Figure 5(a) but now showing effects from three loci. Differences in peak frequency and rate of loss of adaptation reflect a small level of inter-locus interference with selection, where the other loci blunt slightly the efficiency of negative selection on alleles losing their adaptation. Data are extracted from the same simulation used in Figure 4(a).

(c) A clonal population ($N$ = 3000; host mutation rate, $\mu_h$ = 10$^{-4}$ bits/allele/generation; parasite mutation rate, $\mu_p$ = 10$^{-2}$ bits/allele/generation; number of loci, $L$ = 1) at steady state. The frequency peak height is now half of that in Figure 5(a), reflecting capping of allele frequency in clones under substantial lag load. The adaptation score initially declines in a similar manner to (a) but long tails persist in both frequency and adaptation scores. The data are from a single-locus genotype. Figure 5(c) reflects interference between alleles, where poorly adapted alleles are prevented from rapid elimination by coupling with (initially) well-adapted partners.

(d) A clonal population ($N$ = 3000; host mutation rate, $\mu_h$ = $10^{-4}$ bits/allele/generation; parasite mutation rate, $\mu_p$ = $10^{-2}$ bits/allele/generation; number of loci, $L$ = 3) at steady state. In this 3-locus simulation, there is significant inter-locus interference, with lower peak frequency than (c) and much longer tails for losses of allele frequency and adaptation. Data reflect composite time-course data used in the simulation shown in Figure 4(b).

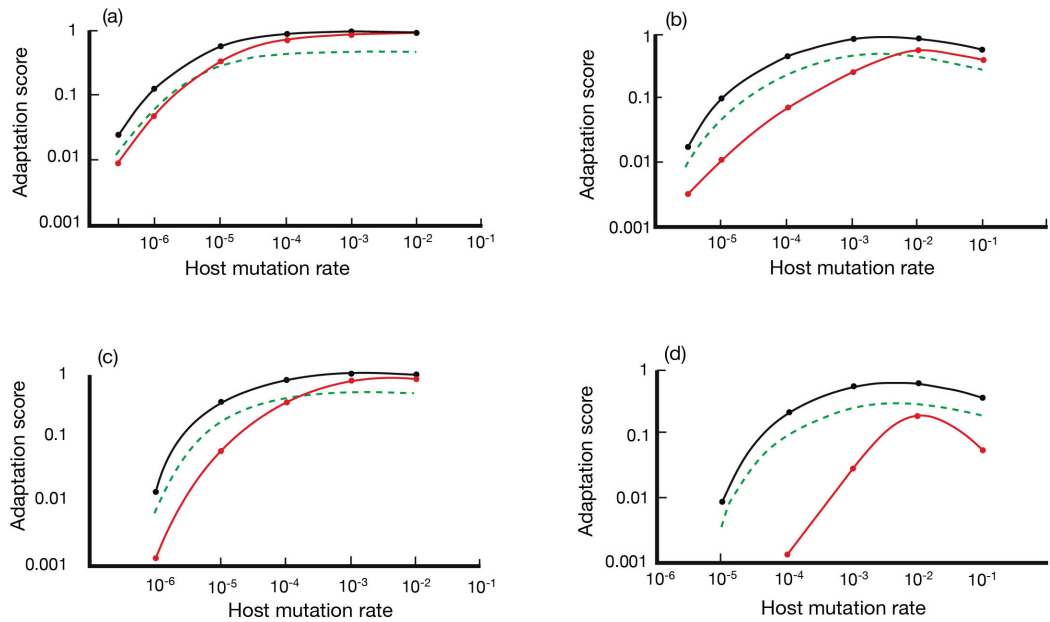

**Figure 6.** Adaptation profiles for sexual and clonal host populations at fixed parasite-mutation rates.

(a) A sexual population ($N$ = 3000; $\mu_p$ = $10^{-4}$ bits/allele/generation; $\mu_h$ = $10^{-7}$-$10^{-2}$ bits/allele/ generation; number of loci, $L$ = 10) (black curve), and a clonal population under the same conditions (red curve) at steady states. The simulation shows that for a parasite mutation rate, $\mu_p$, that is fixed, hosts can optimize their adaptation scores by increasing their own mutation rates until it is similar to or faster than that of the parasite. Adaptation scores differ between sexual and clonal populations at low host mutation rates, where the KJ effect can be expected to be greatest, and approach full adaptation as hosts adopt faster mutation rates. There is a very small mutational load under the conditions of the simulation. Curves of relative fitness (green curve) (= adaptation scores adjusted for fecundity) show that cloning displaces sex as the populations optimize. (b) A sexual population ($N$ = 3000; $\mu_p$ = $10^{-3}$ bits/allele/generation; $\mu_h$ = $10^{-6}$-$10^{-1}$ bits/allele/ generation; number of loci, $L$ = 10) (black curve) and a clonal population under the same conditions (red curve) at steady states. Differences in adaptation scores between clone and sex increase when compared with data in Figure 6(a) above, an effect attributable to the effect of increased host mutation rates on levels of host polymorphism and polyclonality. The increase in polyclonality disproportionately and adversely affects the adaptation of the clonal population. Sexual populations are fitter than clonal ones throughout the range, except at the point of optimal adaptation. (c) A sexual population ($N$ = 3000; $\mu_p$ = $10^{-4}$ bits/allele/generation; $\mu_h$ = $10^{-6}$-$10^{-2}$ bits/allele/ generation; number of loci, $L$ = 20) (black curve), and a clonal population under the same conditions (red curve) at steady states. The shape of the adaptation profiles is similar to that in Figure 6(a), but the larger number of loci in play increases the difference in adaptation at large lag loads. A detectable mutational load is beginning to emerge. (d) Sexual population ($N$ = 3000; $\mu$ = $10^{-3}$ bits/allele/generation; ; $\mu_p$ = $10^{-6}$-$10^{-1}$ bits/allele/ generation; number of loci, $L$ = 20) (black curve) and a clonal population under the same conditions, (red curve) at steady-states. The data are similar to those in Figures 6(a-c) but the combination of a larger number of loci and higher parasite mutation rate markedly suppresses the adaptation of the clonal population. Sexual populations are now fitter over the entire range of host mutation rates.

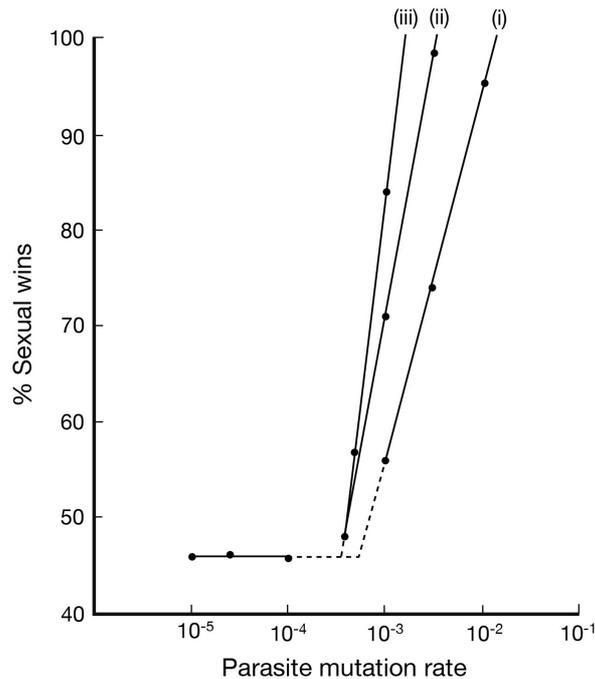

**Figure 7.** Response of sexual populations to injection of clones.

Sexual populations (N = $10^2$ (i), $10^3$ (ii), $10^4$ (iii)) under lag load are allowed to stabilize at steady state rates of parasite and host mutation ($\mu_p/\mu_h$ = 1, $L$ = 10). A constant value for $\mu_p/\mu_h$ is an attempt to generate similar lag loads at all data points. After stabilization, a sexual female of the highest adaptation score is converted to clonality and given a relative fecundity of 2. Because the highest adaptation scores are present only in new advantageous mutations that are still rare, the parasite population develops a negative-frequency dependent response as the frequency of the clonal allele rises. If the parasite mutates rapidly, and the sexual population is under significant lag load, the original host clone and her descendants are progressively rendered unfit and are extinguished by sex. As the parasite mutation rate, $\mu_p$, is reduced in $\mu_p/\mu_h$ (with concomitant reduction in $\mu_h$), the rate of loss of clonal fitness is lower during expansion of the clonal population, and clones outcompete sex. Whether sex wins or clones win depends on population size and $\mu_p$, for constant $\mu_p/\mu_h$. The three lines in Figure 6 show that large populations are more effective at preventing clonal invasion than small ones, as expected. All three populations show threshholds of ~45% for $\mu_p$ below which sex only succeeds because clonal mothers are eliminated by drift immediately after conversion of the sexual mother.